\documentclass[11pt,letterpaper]{article}
\usepackage{soul}
\usepackage{jheppub}
\usepackage[mathletters]{ucs}
\usepackage[utf8]{inputenc}
\usepackage[T1]{fontenc}
\usepackage{amsfonts}
\usepackage{amsmath}
\usepackage{graphicx,subfigure}
\usepackage{tensor} 
\usepackage{mathtools} %
\allowdisplaybreaks[4]
\usepackage{amssymb}
\usepackage{amsthm,amsmath,amssymb}
\usepackage{mathrsfs}
\usepackage{euscript}
\usepackage{color}
\usepackage{braket}
\usepackage{orcidlink}

\title{ \boldmath Dyonic Black Holes in Lorentz-Violating Gravity with a Background Kalb--Ramond Field}

\author[a,b,c]{Yu-Xuan Lin,}
\emailAdd{linyx2023@lzu.edu.cn}

\author[a,b,c]{Jia-Zhou Liu\orcidlink{0009-0005-5430-4258},}
\emailAdd{liujzh2025@lzu.edu.cn}

\author[a,b,c]{Yu-Xiao Liu\orcidlink{0000-0002-4117-4176}\footnote{Corresponding author}}
\emailAdd{liuyx@lzu.edu.cn}

\affiliation[a]{Key Laboratory of Quantum Theory and Applications of MoE, Lanzhou Center for Theoretical Physics, Lanzhou University, Lanzhou 730000, China\vspace{0.1cm}}

\affiliation[b]{ Key Laboratory of Theoretical Physics of Gansu Province, Institute of Theoretical Physics $\&$ Research Center of Gravitation, Lanzhou University, Lanzhou 730000, China \vspace{0.1cm}}

\affiliation[c]{School of Physical Science and Technology, Lanzhou University, Lanzhou 730000, China \vspace{0.1cm}}

\abstract{By introducing a nonminimal coupling between the Kalb--Ramond field and the electromagnetic field, we construct an exact four-dimensional static, spherically symmetric dyonic black hole solution in Lorentz-violating gravity with a background Kalb--Ramond field. The curvature invariants show that the spacetime retains a genuine curvature singularity at $r=0$. We then analyze the geodesic motion of null and timelike particles and obtain the photon-sphere radius, the shadow radius, and the innermost stable circular orbit, demonstrating that both the Lorentz-violating parameter and the dyonic charges can appreciably modify the shadow size and the domain of stable circular motion. In the extended phase space, we derive the thermodynamic quantities and verify the first law of black hole thermodynamics together with the Smarr relation. The system also exhibits a first-order  phase transition between small and large black holes, and its phase structure is strongly influenced by the Lorentz-violating parameter and the dyonic charges.}


\begin{document}
\maketitle
\newpage
\flushbottom

	\section{Introduction}

General relativity and quantum mechanics are the two cornerstones of modern physics, governing gravitational phenomena on macroscopic scales and matter dynamics on microscopic scales, respectively. Despite their remarkable success in their own domains, they have not yet been unified into a self-consistent theoretical framework. To address this tension, a number of approaches to quantum gravity have been developed, including string theory, loop quantum gravity, and holography~\cite{Birrell_Davies_1982,Maldacena_1999,Aharony_2000,Gubser_1998,PhysRevLett.84.2318,PhysRevD.65.103509,ROVELLI199080}. Since the characteristic effects predicted by these theories are generally expected to appear near the Planck scale ($\sim 10^{19}\,\mathrm{GeV}$), they remain far beyond the reach of current experiments and observations. For this reason, the search for possible low-energy signatures of quantum gravity has become an important theme in contemporary theoretical physics. Among the candidate signatures, Lorentz-symmetry breaking is widely regarded as one of the most promising possibilities.

The idea of spontaneous Lorentz-symmetry breaking was first proposed in the context of string theory~\cite{PhysRevD.39.683} and later motivated the construction of the Standard-Model Extension (SME), which provides a systematic effective-field-theory framework for parametrizing Lorentz-violating effects~\cite{PhysRevD.55.6760,PhysRevD.58.116002,PhysRevD.69.105009,RevModPhys.83.11,Tasson_2014,Liberati_2013}. Over the past decades, the SME has driven substantial progress in particle physics, gravitational theory, and precision experimental tests~\cite{PhysRevD.74.045001,PhysRevLett.102.010402,PhysRevD.83.016013,PhysRevD.80.015020,Hees_2016,PhysRevLett.112.111103}. In the gravitational sector, one of the most widely studied toy models is the Bumblebee model, in which a vector field $B_\mu$ couples nonminimally to gravity and acquires a nonzero vacuum expectation value (VEV) through an appropriate potential, thereby inducing spontaneous Lorentz-symmetry breaking in the background spacetime~\cite{PhysRevD.71.065008,Bluhm2008NGmodes,Bertolami_2005}. Within this framework, a variety of exact black hole solutions have been constructed~\cite{PhysRevD.97.104001,PhysRevD.103.044002,Ding_2020,Ding_2021,G_ll__2022,Jha:2020fdj,Ding_2023,Liu_2025,Li_2026,liu2025exactblackholesolutions}, and many of their physical properties have been studied in detail~\cite{Kuang2022StrongLensingKerrLike,Oliveira2019BumblebeeWormholeQNM,Liu2023QNMsSlowlyRotatingEinsteinBumblebee,Gomes2020ThermodynamicsSchwarzschildLike,Kanzi2019GUPModifiedHawkingBumblebee,Oliveira2021QNFrequenciesBumblebee,Kanzi2021GreybodyKerrLikeBumblebee,Ovgun2018LensingWeylBumblebee,Sakalli2023ModifiedHawkingBumblebee,Mangut2023ProbingLIVBumblebee,Uniyal2023ObservableHigherDimBumblebee}.

Besides the Bumblebee scenario, the Kalb--Ramond model provides another important framework for exploring Lorentz-violating gravity. In this model, the rank-two antisymmetric tensor field $B_{\mu\nu}$, which appears as a fundamental excitation in the string spectrum~\cite{PhysRevD.9.2273}, can also acquire a nonzero VEV through nonminimal couplings to gravity together with a suitable potential, thereby triggering spontaneous Lorentz-symmetry breaking. In recent years, black hole and wormhole solutions in Kalb--Ramond gravity have attracted considerable attention. A broad class of exact solutions has been obtained, including Schwarzschild-like black holes, slowly rotating black holes, charged black holes, traversable wormholes, and black holes surrounded by anisotropic fluids~\cite{Liu2024StaticNeutralKR,Yang_2023,Lessa_2020,Sekhmani:2026gup,Duan_2024,sekhmani2026blackholesolutionssurrounded,Liu2025ShadowSlowlyRotatingKR}. Their physical properties have likewise been investigated extensively~\cite{Du2025PhaseStructureKRdS,Guo2024QNMLorentzViolatingBH,Junior2024KRConstraints,Jumaniyozov2024QPOKR,Ma2024SchottkyKRdS,AlBadawi2024GeodesicsChargedKR,Zahid2024ShadowQNMRotatingKR,Zahid2024ElectricPenroseKR,Atamurotov2022ParticleDynamicsKR,Liu2025LorentzViolationEntanglement,Jha2025ObservationalSignatureLV,Fathi2025GlobalMonopoleKR,Filho2024AntisymmetricTensorBH,Filho2025ParticleCreationKR,Hosseinifar2024ShadowsGreybodyKRCharged,Liu2025ShadowSlowlyRotatingKR,Kumar_2020,gu2025quasinormalmodeselectricallycharged}.

In Ref.~\cite{Duan_2024}, the Reissner--Nordstr\"om-like black hole solutions in Kalb--Ramond gravity were derived; however, this framework does not accommodate magnetic charge. In the present work, we introduce a nonminimal coupling between the Kalb--Ramond and electromagnetic fields and allow for both electric and magnetic charges. This setup leads to an exact four-dimensional static, spherically symmetric dyonic black hole solution. We then systematically study its geometric properties, orbital dynamics, and thermodynamic behavior.

The remainder of this paper is organized as follows. In Sec.~\ref{SS2}, we present the theoretical framework of Lorentz-violating gravity with a background Kalb--Ramond field, construct the action containing the nonminimal coupling between the Kalb--Ramond and electromagnetic fields, and derive the corresponding field equations. In Sec.~\ref{SS3}, assuming a static, spherically symmetric metric and a dyonic electromagnetic gauge potential, we simplify the field equations. In Sec.~\ref{SS4}, we obtain exact dyonic black hole solutions in Kalb--Ramond gravity for the two cases $\Lambda=0$ with a quadratic potential and $\Lambda\neq 0$ with a linear potential, and analyze the associated curvature invariants. In Sec.~\ref{SS5}, we study the orbital motion of test particles, including the photon sphere, the black hole shadow, and the innermost stable circular orbit. In Sec.~\ref{SS6}, we discuss black hole thermodynamics and phase transitions in the extended phase space. Finally, Sec.~\ref{SS7} summarizes our conclusions and presents a brief outlook.
\section{Lorentz-Violating Gravity with a Background Kalb--Ramond Field}
\label{SS2}
In this section, we formulate Lorentz-violating gravity in the presence of a background Kalb--Ramond field. We consider the action~\cite{PhysRevD.81.065028}:
\begin{equation}
	S = \int d^4x \sqrt{-g} \left[ \frac{1}{2\kappa} \left( R - 2\Lambda + \xi B_{\mu}{}^{\rho} B_{\nu\rho} R^{\mu\nu} \right) - \frac{1}{12} H_{\mu\nu\rho} H^{\mu\nu\rho} - V(B_{\mu\nu} B^{\mu\nu} \pm b^2) + \mathcal{L}_{\text{matter}} \right].
\end{equation}
Here, $\kappa = \frac{8\pi G}{c^4}$ is the gravitational coupling constant, $\Lambda$ denotes the cosmological constant, and $\xi$ measures the strength of the nonminimal coupling between the Kalb--Ramond field and the Ricci tensor. The potential $V(B_{\mu\nu} B^{\mu\nu} \pm b^2)$ describes the self-interaction of the Kalb--Ramond field and is chosen such that the field can acquire a nonzero VEV through spontaneous symmetry breaking. The field-strength tensor of the Kalb--Ramond field is the totally antisymmetric 3-form
\begin{equation}
	H_{\mu\nu\rho}= \partial_{[\mu} B_{\nu\rho]} = \partial_\mu B_{\nu\rho} + \partial_\nu B_{\rho\mu} + \partial_\rho B_{\mu\nu}.
\end{equation}
which is invariant under the gauge transformation of $B_{\nu\rho}$,
\begin{equation}
	B_{\nu\rho} \to B_{\nu\rho} + \partial_\nu \Lambda_\rho - \partial_\rho \Lambda_\nu,
\end{equation}
where $\Lambda_\rho$ is an arbitrary 1-form field~\cite{PhysRevD.81.065028}. For later convenience, the antisymmetric tensor $B_{\mu\nu}$ can be decomposed as~\cite{PhysRevD.81.065028}
\begin{equation}
	B_{\mu\nu} = \tilde{E}_{[\mu} v_{\nu]} + \epsilon_{\mu\nu\alpha\beta} v^\alpha \tilde{B}^\beta,
\end{equation}
where $v^\mu$ is a timelike four-vector. The background vectors $\tilde{E}_\mu$ and $\tilde{B}_\mu$ may be interpreted as pseudo-electric and pseudo-magnetic fields, respectively. Both are spacelike and satisfy the orthogonality conditions
\begin{equation}
	\tilde{E}_\mu v^\mu = 0, \quad \tilde{B}_\mu v^\mu = 0.
\end{equation}

To study the nonminimal coupling between the Kalb--Ramond field and the electromagnetic field more systematically, we introduce a matter-sector Lagrangian density containing two independent coupling structures:
\begin{equation}
	\mathcal{L}_{\text{matter}} = -\frac{1}{2\kappa} \left(F_{\mu\nu} F^{\mu\nu} +\gamma_1 B^{\mu\nu} B^{\rho\sigma} F_{\mu\nu} F_{\rho\sigma} +\gamma_2 B^{\mu\nu} B_{\mu\nu} F_{\rho\sigma} F^{\rho\sigma}\right).
\end{equation}
Here, $F_{\mu\nu} = \partial_\mu A_\nu - \partial_\nu A_\mu$ is the electromagnetic field-strength tensor. The parameters $\gamma_1$ and $\gamma_2$ denote the coupling constants associated with the tensorial and scalar-trace interactions, respectively. Varying the total action with respect to the metric tensor $g^{\mu\nu}$, the gauge field $A_\mu$, and the Kalb--Ramond field $B_{\mu\nu}$ yields three sets of field equations governing the dynamics of the system.

 First, variation with respect to the metric tensor $g^{\mu\nu}$ yields the modified Einstein field equations:
\begin{equation}
	G_{\mu\nu} + \Lambda g_{\mu\nu} = \kappa \left( T_{\mu\nu}^{\mathrm{KR}} + T_{\mu\nu}^{\mathrm{EM}} \right),
	\label{field equation}
\end{equation}
where $G_{\mu\nu} = R_{\mu\nu} - \frac{1}{2} g_{\mu\nu} R$ is the Einstein tensor. The energy-momentum tensor $T_{\mu\nu}^{\mathrm{KR}}$, which arises from the intrinsic dynamics of the Kalb--Ramond field together with its nonminimal coupling to gravity, takes the form
\begin{align}
	T_{\mu\nu}^{\mathrm{KR}} &= \frac{1}{2} H_{\mu\alpha\beta} H_{\nu}{}^{\alpha\beta} - \frac{1}{12} g_{\mu\nu} H^{\alpha\beta\rho} H_{\alpha\beta\rho} + 4V'(X) B_{\alpha\mu} B^{\alpha}{}_{\nu} - g_{\mu\nu} V(X) \nonumber \\
	&\quad + \frac{\xi}{\kappa} \Bigg[ \frac{1}{2} g_{\mu\nu} B^{\alpha\gamma} B^{\beta}{}_{\gamma} R_{\alpha\beta} - B^{\alpha}{}_{\mu} B^{\beta}{}_{\nu} R_{\alpha\beta} - B^{\alpha\beta} B_{\nu\beta} R_{\mu\alpha} - B^{\alpha\beta} B_{\mu\beta} R_{\nu\alpha} \nonumber \\
	&\quad\quad + \frac{1}{2} \nabla_\alpha \nabla_\mu \left( B^{\alpha\beta} B_{\nu\beta} \right) + \frac{1}{2} \nabla_\alpha \nabla_\nu \left( B^{\alpha\beta} B_{\mu\beta} \right) \nonumber \\
	&\quad\quad - \frac{1}{2} \nabla^\alpha \nabla_\alpha \left( B_\mu{}^\gamma B_{\nu\gamma} \right) - \frac{1}{2} g_{\mu\nu} \nabla_\alpha \nabla_\beta \left( B^{\alpha\gamma} B^{\beta}{}_{\gamma} \right) \Bigg].
\end{align}
Here we define
\begin{equation}
	X = B^{\mu\nu}B_{\mu\nu} \pm b^2, \qquad V'(X) = \frac{dV}{dX}.
\end{equation}
The energy-momentum tensor $T_{\mu\nu}^{\mathrm{EM}}$ contributed by the electromagnetic sector and the two coupling terms is given by
\begin{align}
	T_{\mu\nu}^{\mathrm{EM}} &= \left(1 + \gamma_2 B_{\alpha\beta} B^{\alpha\beta}\right) \left( 2F_{\mu\alpha} F_\nu{}^\alpha - \frac{1}{2} g_{\mu\nu} F_{\alpha\beta} F^{\alpha\beta} \right) + 2\gamma_2 F_{\alpha\beta} F^{\alpha\beta} B_{\mu}{}^{\rho} B_{\nu\rho} \nonumber \\
	&\quad + 2\gamma_1 \left(B_{\alpha\beta} F^{\alpha\beta}\right) \left( B_{\mu}{}^\rho F_{\nu\rho} + B_{\nu}{}^\rho F_{\mu\rho} \right) - \frac{1}{2} \gamma_1 g_{\mu\nu} \left(B_{\alpha\beta} F^{\alpha\beta}\right)^2.
\end{align}
Variation with respect to the gauge field $A_\mu$ yields the Maxwell equations modified by the background Kalb--Ramond field:
\begin{equation}
	\nabla_\nu \left[ \left(1 + \gamma_2 B_{\alpha\beta} B^{\alpha\beta}\right) F^{\mu\nu} + \gamma_1 B_{\alpha\beta} F^{\alpha\beta} B^{\mu\nu} \right] = 0.
	\label{maxwell}
\end{equation}
This equation governs the propagation of the electromagnetic field in a Lorentz-violating background. Relative to the standard Maxwell equations, the deviation encodes the modification of electromagnetic interactions induced by the VEV of the background Kalb--Ramond field. Finally, variation with respect to the Kalb--Ramond field $B_{\mu\nu}$ yields its dynamical equation:
\begin{equation}
	\nabla^\alpha H_{\mu\nu\alpha} - 4 B_{\mu\nu} V'(X) - \frac{2\gamma_1}{\kappa} B_{\alpha\beta} F^{\alpha\beta} F_{\mu\nu} - \frac{2\gamma_2}{\kappa} F_{\alpha\beta} F^{\alpha\beta} B_{\mu\nu} - \frac{\xi}{\kappa} \left( R_{\mu\alpha} B_{\nu}{}^{\alpha} - R_{\nu\alpha} B_{\mu}{}^{\alpha} \right) = 0.
	\label{Bequation}
\end{equation}
This equation shows that the dynamics of the Kalb--Ramond field is controlled not only by its own potential, but also by corrections induced jointly by spacetime curvature and the two electromagnetic coupling terms.
\section{Metric Ansatz and the Field Equations}\label{SS3}

For a static, spherically symmetric spacetime, we take the metric ansatz as
\begin{equation}
	ds^2 = -F(r)dt^2 + G(r)^{-1}dr^2 + r^2(d\theta^2 + \sin^2\theta d\phi^2),
\end{equation}
where $F(r)$ and $G(r)$ are two unknown functions depending only on the radial coordinate $r$, characterizing the temporal and radial components of the spacetime geometry, respectively.

Through spontaneous Lorentz-symmetry breaking, the Kalb--Ramond field acquires a background VEV with constant modulus at the minimum of the potential. Assuming that $X=0$ at this point, one has
\begin{equation}
	B_{\mu\nu} B^{\mu\nu} = \mp b^2.
\end{equation}
To obtain a vacuum configuration compatible with static spherical symmetry, we choose a ``pseudo-electric'' background configuration containing only the time-radial components:
\begin{equation}
	B_{\mu\nu} =
	\begin{pmatrix}
		0 & \frac{b}{\sqrt{2}}\sqrt{F(r)G(r)} & 0 & 0 \\
		-\frac{b}{\sqrt{2}}\sqrt{F(r)G(r)} & 0 & 0 & 0 \\
		0 & 0 & 0 & 0 \\
		0 & 0 & 0 & 0
	\end{pmatrix}.
\end{equation}
It is straightforward to verify that, for this configuration,
\begin{equation}
	B_{\mu\nu} B^{\mu\nu} = -b^2,
\end{equation}
and that the field strength tensor of the Kalb--Ramond field vanishes identically, namely,
\begin{equation}
	H_{\mu\nu\rho} = 0.
\end{equation}

For the electromagnetic field, we adopt a gauge-potential ansatz containing both a radial scalar potential $\Phi(r)$ and a magnetic monopole charge $p$:
\begin{equation}
	A_{\mu} = \left( -\Phi(r), 0, 0, p \cos\theta \right).
\end{equation}
This configuration describes a static, spherically symmetric electromagnetic system carrying both electric charge $Q$ and magnetic charge $p$. The corresponding nonvanishing components of the electromagnetic field tensor are
\begin{equation}
	F_{tr}=-F_{rt} = -\Phi'(r), \qquad F_{\theta\phi} =-F_{\phi\theta}= -p \sin\theta.
\end{equation}

Substituting the metric ansatz, the vacuum configuration of the Kalb--Ramond field, and the electromagnetic potential into Eq.~\eqref{field equation}, we obtain a system of ordinary differential equations for the metric functions $F(r)$ and $G(r)$ together with the electric potential $\Phi(r)$. The three independent nonvanishing components of the modified Einstein equations are
\begin{align}
	-\frac{G'}{r G^2} - \frac{1}{r^2} + \frac{1}{r^2 G} + \Lambda
	&= \frac{b^2 \xi F'^2}{4 F^2 G} + \frac{b^2 \xi F' G'}{4 F G^2} - \frac{b^2 \xi F''}{2 F G} - \frac{b^2 \xi F'}{2 r F G} - \frac{\Phi'^2}{F G} \nonumber \\
	&\quad + \frac{3 b^2 (\gamma_1+\gamma_2) \Phi'^2}{F G} - \frac{b^2 \gamma_2 p^2}{r^4} - \frac{p^2}{r^4} - 2 b^2 V' \kappa + \frac{b^2 \xi}{2 r^2 G},
	\label{eq:einstein_tt}
\end{align}
\begin{align}
	\frac{F'}{r F G} - \frac{1}{r^2} + \frac{1}{r^2 G} + \Lambda
	&= \frac{b^2 \xi F'^2}{4 F^2 G} + \frac{b^2 \xi F' G'}{4 F G^2} - \frac{b^2 \xi F''}{2 F G} + \frac{b^2 \xi G'}{2 r G^2} - \frac{\Phi'^2}{F G} \nonumber \\
	&\quad + \frac{3 b^2 (\gamma_1+\gamma_2) \Phi'^2}{F G} - \frac{b^2 \gamma_2 p^2}{r^4} - \frac{p^2}{r^4} - 2 b^2 V' \kappa + \frac{b^2 \xi}{2 r^2 G},
	\label{eq:einstein_rr}
\end{align}
\begin{align}
	\Lambda - \frac{F'^2}{4 F^2 G} - \frac{F' G'}{4 F G^2} + \frac{F''}{2 F G} + \frac{F'}{2 r F G} - \frac{G'}{2 r G^2}
	&= -\frac{b^2 \xi F'^2}{8 F^2 G} - \frac{b^2 \xi F' G'}{8 F G^2} + \frac{b^2 \xi F''}{4 F G} + \frac{b^2 \xi F'}{4 r F G} \nonumber \\
	&\quad - \frac{b^2 \xi G'}{4 r G^2} - \frac{b^2 \left(\gamma_1+\gamma_2\right) \Phi'^2}{F G} + \frac{\Phi'^2}{F G} \nonumber \\
	&\quad - \frac{b^2 \gamma_2 p^2}{r^4} + \frac{p^2}{r^4}.
	\label{eq:einstein_theta}
\end{align}
From Eqs.~\eqref{maxwell} and \eqref{Bequation}, we further obtain the modified Maxwell equation and the equation of motion for the Kalb--Ramond field in the forms
\begin{align}
	\frac{b \xi F'^2}{4 \sqrt{2} \kappa F^{3/2} G^{3/2}}
	+ \frac{b \xi F' G'}{4 \sqrt{2} \kappa F^{1/2} G^{5/2}}
	- \frac{b \xi F''}{2 \sqrt{2} \kappa F^{1/2} G^{3/2}}
	- \frac{b \xi F'}{2 \sqrt{2} \kappa r F^{1/2} G^{3/2}} \nonumber \\
	+ \frac{\sqrt{2} b \left(\gamma_1+\gamma_2\right) \Phi'^2}{\kappa F^{1/2} G^{3/2}}
	+ \frac{b \xi F^{1/2} G'}{2 \sqrt{2} \kappa r G^{5/2}}
	- \frac{\sqrt{2} b \gamma_2 p^2 F^{1/2}}{\kappa r^4 G^{1/2}}
	- \frac{\sqrt{2} b V' F^{1/2}}{G^{1/2}} &= 0,
	\label{eq:kr_eq}
\end{align}
\begin{align}
	\frac{F' \Phi'}{\kappa F G}
	- \frac{b^2 (\gamma_1 + \gamma_2) F' \Phi'}{\kappa F G}
	+ \frac{G' \Phi'}{\kappa G^2}
	- \frac{b^2 (\gamma_1 + \gamma_2) G' \Phi'}{\kappa G^2} \nonumber \\
	- \frac{4 \Phi'}{\kappa r G}
	+ \frac{4 b^2 (\gamma_1 + \gamma_2) \Phi'}{\kappa r G}
	- \frac{2 \Phi''}{\kappa G}
	+ \frac{2 b^2 (\gamma_1 + \gamma_2) \Phi''}{\kappa G} &= 0.
	\label{eq:maxwell_eq}
\end{align}
A further comparison of the first two components of the gravitational field equations, namely Eqs.~\eqref{eq:einstein_tt} and \eqref{eq:einstein_rr}, shows that their right-hand sides are identical. Subtracting the two equations, the left-hand side reduces to
\begin{equation}
	-\frac{G'}{r G^2} - \frac{F'}{r F G} = 0.
\end{equation}
This can be rearranged as
\begin{equation}
	\frac{F'(r)}{F(r)} + \frac{G'(r)}{G(r)} = 0.
\end{equation}
Integrating the above equation, one obtains
\begin{equation}
	F(r)G(r) = C,
\end{equation}
where $C$ is an integration constant. By a rescaling of the time coordinate without loss of generality, this constant can be set to unity, namely,
\begin{equation}
	G(r) = \frac{1}{F(r)}.
\end{equation}
This relation greatly simplifies the subsequent analysis.

Substituting $G(r)=1/F(r)$ into Eqs.~\eqref{eq:einstein_tt}--\eqref{eq:maxwell_eq} and performing straightforward algebraic simplifications, the original system can be reduced to the following four independent ordinary differential equations:
\begin{align}
	\frac{r F' + F - 1}{r^2} + \Lambda
	&= \frac{1}{2} \left[ -b^2 \xi F'' - \frac{b^2 \xi F'}{r} + \frac{b^2 \xi F}{r^2} - \frac{2 b^2 \gamma_2 p^2}{r^4} \right. \nonumber \\
	&\qquad \left. + 6 b^2 \left(\gamma_1+\gamma_2\right) \Phi'^2- 4 b^2 V' \kappa - \frac{2 p^2}{r^4} - 2 \Phi'^2 \right],
	\label{eq:simplified_1}
\end{align}
\begin{align}
	\frac{1}{2} F'' + \frac{F'}{r} + \Lambda
	&= \frac{1}{4} b^2 \xi F'' + \frac{b^2 \xi F'}{2 r} - \frac{b^2 \gamma_2 p^2}{r^4} - \left[ b^2 (\gamma_1 + \gamma_2) - 1 \right]\Phi'^2 + \frac{p^2}{r^4},
	\label{eq:simplified_2}
\end{align}
\begin{align}
	\frac{b \xi F F''}{2 \sqrt{2} \kappa}
	+ \frac{b \xi F F'}{\sqrt{2} \kappa r}
	+ \frac{\sqrt{2} b \gamma_2 p^2 F}{\kappa r^4}
	- \frac{\sqrt{2} b \left(\gamma_1+\gamma_2\right) F \Phi'^2}{\kappa}
	+ \sqrt{2} b F V' &= 0,
	\label{eq:simplified_3}
\end{align}
\begin{align}
	\frac{2 F \left[ b^2 (\gamma_1 + \gamma_2) - 1 \right] \left( r \Phi'' + 2 \Phi' \right)}{\kappa r} &= 0.
	\label{eq:simplified_4}
\end{align}

In the next section, we derive the corresponding exact black hole solutions for different choices of the potential and different cosmological-constant backgrounds.
\section{Exact Analytical Solutions for Electromagnetically Dyonic Black Holes}\label{SS4}

In this section, we solve the reduced system obtained in the previous section for two physically representative cases. The first corresponds to a vanishing cosmological constant with a quadratic self-interaction potential for the Kalb--Ramond field, whereas the second corresponds to a nonvanishing cosmological constant with a linear potential.

\subsection{Case 1: $\Lambda=0$ and $V(X)=\frac{1}{2}\lambda X^2$}

We first consider an asymptotically flat background with vanishing cosmological constant and assume that the self-interaction potential of the Kalb--Ramond field takes the quadratic form~\cite{Bluhm_2008}
\begin{equation}
	V(X) = \frac{1}{2}\lambda X^2,
\end{equation}
where $\lambda$ is a Lagrange multiplier field. Variation with respect to $\lambda$ yields the constraint $X=0$, which ensures that the potential is minimized, corresponding to $B_{\mu\nu}B^{\mu\nu}=-b^2$. At this minimum, one has
\begin{equation}
	V(X)=0, \qquad V'(X)=0,
\end{equation}
so that the potential term no longer contributes to the field equations.

\subsubsection{Purely electric black hole solution ($p=0$)}

As the simplest case, we first consider a purely electric black hole with vanishing magnetic charge. Setting $p=0$, the system of Eqs.~\eqref{eq:simplified_1}--\eqref{eq:simplified_4} simplifies further. In particular, Eq.~\eqref{eq:simplified_4} implies that, provided $b^2(\gamma_1+\gamma_2)\neq 1$, one must have
\begin{equation}
	r\Phi''+2\Phi'=0.
\end{equation}
The general solution of this equation is
\begin{equation}
	\Phi(r)=-\frac{c_1}{r}+c_3,
\end{equation}
where $c_1$ and $c_3$ are integration constants.

Substituting this solution for the electric potential into the remaining three equations, Eqs.~\eqref{eq:simplified_1}--\eqref{eq:simplified_3}, and solving them simultaneously, one finds that the system admits a nontrivial self-consistent solution only when the coupling constants satisfy
\begin{equation}
	\gamma_1+\gamma_2=\frac{\xi}{2}.
\end{equation}
Under this condition, the general solution for the metric function $F(r)$ can be written as
\begin{equation}
	F(r)=\frac{1}{1-\ell}+\frac{c_2}{r}+\frac{c_1^2}{r^2},
\end{equation}
where we have introduced the dimensionless Lorentz-violating parameter
\begin{equation}
	\ell \equiv \frac{\xi b^2}{2}.
\end{equation}
This parameter encodes both the magnitude of the VEV $b$ of the Kalb--Ramond field and the coupling strength $\xi$ between the Kalb--Ramond field and gravity, and therefore serves as the key quantity characterizing the strength of Lorentz-violating effects. In Ref.~\cite{Yang_2023KR}, experimental data were used to constrain $\ell$ to the range $-1.1 \times 10^{-10} \leq \ell \leq 5.4 \times 10^{-10}$.

To clarify the physical meaning of the integration constant $c_1$, one must account for the modification of Gauss's law induced by the background Kalb--Ramond field. Since the Maxwell equations in a Lorentz-violating background take the form given in Eq.~\eqref{maxwell}, the associated conserved current is no longer simply $\nabla_\nu F^{\mu\nu}$, but should instead be defined as
\begin{equation}
	J^\mu=\nabla_\nu \left[ \left(1+\gamma_2 B_{\alpha\beta} B^{\alpha\beta}\right)F^{\mu\nu}+\gamma_1B_{\alpha\beta} F^{\alpha\beta}B^{\mu\nu} \right].
\end{equation}
The physical electric charge $Q$ is then obtained by integrating this conserved current over a three-dimensional spacelike hypersurface $\Sigma$, and, by Stokes' theorem, converting it into a surface integral over its boundary $\partial\Sigma$~\cite{Carroll2019}:
\begin{align}
	Q &= -\frac{1}{4\pi}\int_\Sigma d x^3 \sqrt{\gamma^{(3)}}\, n_\mu J^\mu \nonumber \\
	&= -\frac{1}{4\pi}\int_{\partial\Sigma} d\theta d\phi \sqrt{\gamma^{(2)}}\, n_\mu \sigma_\nu
	\left[ \left(1+\gamma_2 B_{\alpha\beta} B^{\alpha\beta}\right)F^{\mu\nu}+\gamma_1B_{\alpha\beta} F^{\alpha\beta}B^{\mu\nu} \right] \nonumber \\
	&= c_1 \left[ 1-(\gamma_1+\gamma_2)b^2 \right]
	= c_1 \left( 1-\frac{\xi b^2}{2} \right)
	= c_1(1-\ell).
\end{align}
Here, $\Sigma$ denotes a three-dimensional spacelike region with induced metric $\gamma_{ij}^{(3)}$, while $\partial\Sigma$ is the two-sphere at spatial infinity, whose induced metric is
$\gamma_{ij}^{(2)} = r^2(d\theta^2+\sin^2\theta\, d\phi^2)$. Correspondingly, $n_\mu=(1,0,0,0)$ is the unit normal vector associated with $\Sigma$, and $\sigma_\mu=(0,1,0,0)$ is the unit normal vector associated with $\partial\Sigma$. It follows that
\begin{equation}
	c_1=\frac{Q}{1-\ell},
\end{equation}
so that the electrostatic potential becomes
\begin{equation}
	\Phi(r)=-\frac{Q}{(1-\ell)r}.
\end{equation}
This result shows that the presence of the background Kalb--Ramond field induces a rescaling relation between the physical charge and the integration constant, controlled by the Lorentz-violating parameter $\ell$. In the limit $\ell\to 0$, where Lorentz violation disappears, the standard result $c_1=Q$ is naturally recovered.

We now determine the remaining integration constants. Choosing the electric potential to vanish at spatial infinity, we set $c_3=0$. To ensure that the metric reduces to the standard Reissner--Nordstr\"om (RN) solution in the limit $\ell \to 0$, we set $c_2 = -2M$. The exact purely electric Kalb--Ramond black hole solution can therefore be written as
\begin{align}
	F(r) &= \frac{1}{1-\ell}-\frac{2M}{r}+\frac{Q^2}{(1-\ell)^2 r^2}, \\
	\Phi(r) &= -\frac{Q}{(1-\ell)r}.
\end{align}
Furthermore, as $r \to \infty$, the metric function approaches $F(r) \to \frac{1}{1-\ell} \neq 1$, explicitly indicating that the spacetime is not asymptotically Minkowski. As can be seen, in the purely electric case, our result exactly reproduces the form derived in Ref.~\cite{Duan_2024}.

\subsubsection{Dyonic black hole solution ($p\neq 0$)}

We now turn to the more general case in which the black hole carries both electric charge $Q$ and magnetic charge $p$. When $p\neq 0$, additional $1/r^4$ terms appear in Eqs.~\eqref{eq:simplified_1}--\eqref{eq:simplified_4}. Following the same procedure as in the purely electric case, one finds that the electric potential retains the form
\begin{equation}
	\Phi(r)=-\frac{Q}{(1-\ell)r}.
\end{equation}
However, due to the presence of the magnetic-charge term, the coupling constants must satisfy more restrictive constraints. In addition to
\begin{equation}
	\gamma_1+\gamma_2=\frac{\xi}{2},
\end{equation}
one must also impose
\begin{equation}
	\gamma_2=\frac{\xi}{2b^2\xi-2}.
\end{equation}
These two conditions jointly determine the relation between the electromagnetic coupling constants $\gamma_1$, $\gamma_2$ and the gravitational coupling constant $\xi$. Under these constraints, the exact solution for the metric function is
\begin{equation}
	F(r)=\frac{1}{1-\ell}-\frac{2M}{r}+\frac{Q^2}{(1-\ell)^2 r^2}+\frac{p^2}{(1-2\ell)r^2}.
\end{equation}
This is the static, spherically symmetric black hole solution in Kalb--Ramond gravity carrying both electric and magnetic charges. As in the electric sector, the magnetic charge contributes through a $1/r^2$ correction, although its coefficient is renormalized differently by the Lorentz-violating parameter $\ell$. When $p=0$, the solution reduces smoothly to the purely electric case obtained above.

To clarify the intrinsic geometrical properties and singularity structure of this dyonic Kalb--Ramond black hole, we calculate three representative curvature invariants. These quantities characterize the local strength of the gravitational field and, at the same time, allow one to distinguish coordinate singularities from genuine physical singularities, thereby testing the geometrical completeness of the solution.

First, the Ricci scalar is
\begin{equation}
	R=R_{\mu\nu}g^{\mu\nu}=\frac{2\ell}{(-1+\ell)r^2}.
	\label{Ricci}
\end{equation}
Second, the square of the Ricci tensor is
\begin{equation}
	\begin{aligned}
		R_{\mu\nu}R^{\mu\nu} &= \frac{4p^4}{(1-2\ell)^2 r^8} + \frac{8p^2 Q^2}{(1-2\ell)(-1+\ell)^2 r^8}+ \frac{4Q^4}{(-1+\ell)^4 r^8}  \\
		&\quad + \frac{4p^2 \ell}{(1-2\ell)(-1+\ell)r^6} + \frac{4Q^2 \ell}{(-1+\ell)^3 r^6} + \frac{2\ell^2}{(-1+\ell)^2 r^4}.
	\end{aligned}
	\label{RR}
\end{equation}
Finally, the Kretschmann scalar is
\begin{equation}
	\begin{aligned}
		R_{\mu\nu\rho\lambda}R^{\mu\nu\rho\lambda} &= \frac{56p^4}{(1-2\ell)^2 r^8} + \frac{112p^2 Q^2}{(1-2\ell)(-1+\ell)^2 r^8}+ \frac{56Q^4}{(-1+\ell)^4 r^8}  - \frac{96M p^2}{(1-2\ell)r^7}- \frac{96M Q^2}{(-1+\ell)^2 r^7}  \\
		&\quad - \frac{8p^2 \ell}{(1-2\ell)(-1+\ell)r^6} - \frac{8Q^2 \ell}{(-1+\ell)^3 r^6} + \frac{48M^2}{r^6}+\frac{16M \ell}{(-1+\ell)r^5} + \frac{4\ell^2}{(-1+\ell)^2 r^4}.
	\end{aligned}
	\label{Kretschmann}
\end{equation}
From the above expression, it can be seen that terms of orders  \(r^{-2}\), \(r^{-4}\), \(r^{-5}\), and some \(r^{-6}\) appear only when the Lorentz-violating parameter \(\ell \neq 0\), reflecting the nontrivial corrections to the black hole background geometry induced by Lorentz symmetry breaking. Meanwhile, the extremely high-order terms containing \(r^{-7}\) and \(r^{-8}\) emerge only when the electric charge $Q$ or magnetic charge $p$ is nonzero, and these higher-order terms become more significant as $Q$ and $p$ increase. Most crucially, irrespective of the presence of electromagnetic charges or Lorentz symmetry breaking, \(r=0\) is always a singularity of the Kretschmann scalar. This definitively demonstrates the existence of an absolute and irremovable intrinsic physical singularity at $r=0$.

\subsection{Case 2: $\Lambda\neq 0$ and $V(X)=\lambda X$}

When a nonvanishing cosmological constant is included, the system no longer admits a self-consistent analytical solution if one retains the quadratic potential
$V(X)=\frac{1}{2}\lambda X^2$
and imposes $V'(X)=0$. This feature is consistent with related results previously found in Bumblebee gravity and earlier studies of Kalb--Ramond gravity~\cite{PhysRevD.103.044002,Duan_2024}. Therefore, to obtain analytical solutions in the presence of a nonzero cosmological constant, one must relax the vacuum condition appropriately and instead choose the linear self-interaction potential~\cite{Bluhm_2008}
\begin{equation}
	V(X)=\lambda X.
\end{equation}
In this case, the first derivative of the potential satisfies $V'(X)=\lambda\neq 0$, and enters the field equations as an undetermined constant. Repeating the above solution procedure, one finds that the coupling constants must still satisfy the same constraint conditions as in the $\Lambda=0$ case:
\begin{equation}
	\gamma_1+\gamma_2=\frac{\xi}{2}, \qquad \gamma_2=\frac{\xi}{2b^2\xi-2}.
\end{equation}
At the same time, the Lagrange multiplier $\lambda$ is uniquely determined as
\begin{equation}
	\lambda=\frac{\ell\Lambda}{(\ell-1)b^2\kappa}.
\end{equation}
One then obtains the exact dyonic Kalb--Ramond black hole solution in the presence of a cosmological constant:
\begin{equation}
	F(r)=\frac{1}{1-\ell}-\frac{2M}{r}+\frac{Q^2}{(1-\ell)^2 r^2}+\frac{p^2}{(1-2\ell)r^2}+\frac{\Lambda}{3(\ell-1)}r^2.
\end{equation}
This result shows that the Lorentz-violating parameter $\ell$ also rescales the cosmological-constant term.

In the presence of a cosmological constant, we can calculate the three representative curvature invariants again. First, the Ricci scalar is
\begin{equation}
	R=R_{\mu\nu}g^{\mu\nu}=\frac{2\ell}{(-1+\ell)r^2}-\frac{4\Lambda}{-1+\ell}.
\end{equation}
Second, the square of the Ricci tensor is
\begin{equation}
	\begin{aligned}
		R_{\mu\nu}R^{\mu\nu} &= \frac{4p^4}{(1-2\ell)^2 r^8} + \frac{8p^2 Q^2}{(1-2\ell)(-1+\ell)^2 r^8}+ \frac{4Q^4}{(-1+\ell)^4 r^8} + \frac{4p^2 \ell}{(1-2\ell)(-1+\ell)r^6} \\
		&\quad  + \frac{4Q^2 \ell}{(-1+\ell)^3 r^6} + \frac{2\ell^2}{(-1+\ell)^2 r^4} - \frac{4\ell \Lambda}{(-1+\ell)^2 r^2} + \frac{4\Lambda^2}{(-1+\ell)^2}.
	\end{aligned}
\end{equation}
Finally, the Kretschmann scalar is
\begin{equation}
	\begin{aligned}
		R_{\mu\nu\rho\lambda}R^{\mu\nu\rho\lambda} &= \frac{56p^4}{(1-2\ell)^2 r^8} + \frac{112p^2 Q^2}{(1-2\ell)(-1+\ell)^2 r^8}+ \frac{56Q^4}{(-1+\ell)^4 r^8} - \frac{96M p^2}{(1-2\ell)r^7}  \\
		&\quad  - \frac{96M Q^2}{(-1+\ell)^2 r^7}- \frac{8p^2 \ell}{(1-2\ell)(-1+\ell)r^6} - \frac{8Q^2 \ell}{(-1+\ell)^3 r^6} + \frac{48M^2}{r^6} \\
		&\quad + \frac{16M \ell}{(-1+\ell)r^5} + \frac{4\ell^2}{(-1+\ell)^2 r^4} - \frac{8\ell \Lambda}{3(-1+\ell)^2 r^2} + \frac{8\Lambda^2}{3(-1+\ell)^2}.
	\end{aligned}
\end{equation}
As can be seen, compared with the expressions given in Eqs.~\eqref{Ricci}--\eqref{Kretschmann} for the case without a cosmological constant, the presence of the cosmological constant introduces an additional constant term and an  $r^{-2}$ term. These terms depend on the cosmological constant  $\Lambda$ and the Lorentz-violating parameter $\ell$, and notably, the new $r^{-2}$ term arises only when the Lorentz-violating parameter \(\ell \neq 0\).

\section{Orbital Motion of Test Particles}\label{SS5}
In this section, we analyze how Lorentz violation, the electric charge $Q$, and the magnetic charge $p$ affect geodesic motion in the dyonic Kalb--Ramond black hole background.

The dynamics of a test particle is governed by the Lagrangian
\begin{equation}
	\mathcal{L} = \frac{1}{2} g_{\mu\nu} \dot{x}^\mu \dot{x}^\nu
	= \frac{1}{2} \left[ -F(r)\dot{t}^2 + \frac{1}{F(r)}\dot{r}^2 + r^2\dot{\theta}^2 + r^2\sin^2\theta \dot{\phi}^2 \right],
	\label{eq:lagrangian}
\end{equation}
where an overdot denotes differentiation with respect to the affine parameter $\lambda$ of the particle. Owing to the spherical symmetry of the spacetime, the particle motion is necessarily confined to a fixed plane. Without loss of generality, we choose the equatorial plane $\theta=\pi/2$, so that $\dot{\theta}=0$.

A static, spherically symmetric spacetime admits two Killing vector fields, $\partial_t$ and $\partial_\phi$, associated with time-translation symmetry and axial symmetry, respectively. From the Euler--Lagrange equations, one obtains two conserved quantities, namely the energy per unit mass $E$ and the angular momentum per unit mass $L$:
\begin{align}
	p_t &= \frac{\partial \mathcal{L}}{\partial \dot{t}} = -F(r)\dot{t} \equiv -E,
	\label{eq:energy_conserv} \\
	p_\phi &= \frac{\partial \mathcal{L}}{\partial \dot{\phi}} = r^2\dot{\phi} \equiv L.
	\label{eq:angular_conserv}
\end{align}
In addition, the normalization condition for the four-velocity is
\begin{equation}
	g_{\mu\nu}\dot{x}^\mu \dot{x}^\nu = -\epsilon,
\end{equation}
where $\epsilon=0$ corresponds to a massless photon and $\epsilon=1$ to a massive test particle. Substituting Eqs.~\eqref{eq:energy_conserv} and \eqref{eq:angular_conserv} into the normalization condition, one obtains the radial equation of motion
\begin{equation}
	-F(r)\left(\frac{E}{F(r)}\right)^2 + \frac{1}{F(r)}\dot{r}^2 + r^2\left(\frac{L}{r^2}\right)^2 = -\epsilon.
\end{equation}
This equation can be recast into the Newtonian-like radial energy form
\begin{equation}
	\dot{r}^2 + V_{\text{eff}}^2(r) = E^2, \qquad
	V_{\text{eff}}(r) \equiv \sqrt{F(r)\left(\epsilon + \frac{L^2}{r^2}\right)},
	\label{eq:eff_potential}
\end{equation}
where $V_{\text{eff}}(r)$ is the effective potential governing the radial dynamics of the particle. Here we take the metric function in the general form containing the electric charge $Q$, magnetic charge $p$, and cosmological constant $\Lambda$:
\begin{equation}
	F(r) = \frac{1}{1-\ell} - \frac{2M}{r} + \frac{Q^2}{(1-\ell)^2 r^2} + \frac{p^2}{(1-2\ell)r^2} - \frac{\Lambda r^2}{3(1-\ell)}.
	\label{eq:F_full}
\end{equation}
\subsection{Photon Sphere and Shadow of Massless Photons}

For photons ($\epsilon=0$), the effective potential reduces to
\begin{equation}
	V_{\text{eff}}(r)=\frac{L\sqrt{F(r)}}{r}.
\end{equation}
The unstable circular photon orbit outside the black hole corresponds to the photon sphere, whose radius is determined by the conditions $V_{\text{eff}}=E$ and $\partial_r V_{\text{eff}}=0$. Differentiating the effective potential and setting the result to zero, one obtains the fundamental equation satisfied by the radius of the photon sphere,
\begin{equation}
	2F(r_{\text{ph}})-r_{\text{ph}}F'(r_{\text{ph}})=0.
	\label{eq:photon_sphere}
\end{equation}
Substituting Eq.~\eqref{eq:F_full} into the above equation, one finds that the terms involving $\Lambda$ cancel out exactly, so that the radius of the photon sphere depends only on $M$, $Q$, $p$, and $\ell$, and satisfies
\begin{equation}
	\frac{2}{1-\ell} - \frac{6M}{r_{\text{ph}}}
	+ \frac{4}{r_{\text{ph}}^2}\left( \frac{Q^2}{(1-\ell)^2} + \frac{p^2}{1-2\ell} \right) = 0.
\end{equation}
Taking the larger physically relevant positive root, one obtains
\begin{equation}
	r_{\text{ph}} = \frac{3M(1-\ell)}{2}
	\left[
	1 + \sqrt{
		1 - \frac{8}{9M^2(1-\ell)}
		\left( \frac{Q^2}{(1-\ell)^2} + \frac{p^2}{1-2\ell} \right)
	}
	\right].
	\label{eq:rph_explicit}
\end{equation}

For a static observer located at radial coordinate $r_o$, the critical impact parameter determined by the photon sphere gives the apparent radius of the black hole shadow,
\begin{equation}
	r_{\text{sh}}=\sqrt{\frac{F(r_o)}{F(r_{\text{ph}})}}r_{\text{ph}}.
	\label{eq:shadow_radius}
\end{equation}
Accordingly, this can be written more explicitly as
\begin{equation}
	r_{\mathrm{sh}}
	=
	\sqrt{
		\frac{
			\frac{1}{1-\ell}
			-\frac{2M}{r_{\mathrm{o}}}
			+\frac{Q^2}{(1-\ell)^2r_{\mathrm{o}}^2}
			+\frac{p^2}{(1-2\ell)r_{\mathrm{o}}^2}
			+\frac{\Lambda}{3(\ell-1)}r_{\mathrm{o}}^2
		}{
			\frac{1}{1-\ell}
			-\frac{2M}{r_{\mathrm{ph}}}
			+\frac{Q^2}{(1-\ell)^2r_{\mathrm{ph}}^2}
			+\frac{p^2}{(1-2\ell)r_{\mathrm{ph}}^2}
			+\frac{\Lambda}{3(\ell-1)}r_{\mathrm{ph}}^2
		}
	}\, r_{\mathrm{ph}}.
	\label{eq:shadow_full}
\end{equation}
We now specialize to the $\Lambda=0$ case. If the observer is located at spatial infinity, namely $r_o\to\infty$, then $F(r_o)\to 1/(1-\ell)$. The shadow radius simplifies to
\begin{equation}
	r_{\text{sh}} = \frac{3M(1-\ell)}{2}(1+\Omega)
	\sqrt{\frac{6(1+\Omega)}{1+3\Omega}},
	\label{eq:rsh_simplified}
\end{equation}
where
\begin{equation}
	\Omega =
	\sqrt{
		1 - \frac{8}{9M^2(1-\ell)}
		\left( \frac{Q^2}{(1-\ell)^2} + \frac{p^2}{1-2\ell} \right)
	}.
	\label{eq:omega_def}
\end{equation}
It is evident that the square root in Eq.~\eqref{eq:rph_explicit} must be real. Therefore, the system parameters must satisfy
\begin{equation}
	M^2 \ge \frac{8}{9(1-\ell)}
	\left(
	\frac{Q^2}{(1-\ell)^2} + \frac{p^2}{1-2\ell}
	\right).
	\label{eq:photon_sphere_existence}
\end{equation}
This is the necessary condition for the existence of the photon sphere. When the inequality is saturated, the photon-sphere radius shrinks to its theoretical minimum value,
\begin{equation}
	r_{\text{ph}}^{\min}=\frac{3}{2}M(1-\ell).
\end{equation}
Near this critical point, the event horizon may already have disappeared, while the photon sphere can still survive within a certain parameter range. Only when the parameters deviate further until  Eq.~\eqref{eq:photon_sphere_existence} is no longer satisfied does the photon sphere disappear altogether, and the corresponding black hole shadow and strong-lensing features cease to exist.

To distinguish more clearly the condition for the existence of the photon sphere from that for the event horizon, let us again consider the $\Lambda=0$ case, for which the metric function reduces to
\begin{equation}
	F(r)=\frac{1}{1-\ell}-\frac{2M}{r}
	+\frac{1}{r^2}\left(
	\frac{Q^2}{(1-\ell)^2}+\frac{p^2}{1-2\ell}
	\right).
\end{equation}
The radius of the event horizon $r_+$ is determined by $F(r_+)=0$, which leads to the quadratic equation
\begin{equation}
	\frac{1}{1-\ell}r_+^2 - 2Mr_+
	+\left(
	\frac{Q^2}{(1-\ell)^2}+\frac{p^2}{1-2\ell}
	\right)=0.
\end{equation}
Requiring this equation to admit real roots implies that its discriminant must be nonnegative, which yields the condition for the existence of the event horizon,
\begin{equation}
	M^2 \ge \frac{1}{1-\ell}
	\left(
	\frac{Q^2}{(1-\ell)^2}+\frac{p^2}{1-2\ell}
	\right).
	\label{eq:horizon_existence}
\end{equation}
Comparing Eqs.~\eqref{eq:photon_sphere_existence} and \eqref{eq:horizon_existence}, one sees that, since $8/9<1$, the parameter region supporting a photon sphere is broader than that supporting an event horizon. This implies that, as the electromagnetic charges increase, the event horizon may disappear before the photon sphere does, giving rise to naked singularity spacetimes that still retain a photon sphere. Only when the parameters increase further beyond the existence bound for the photon sphere do the strong lensing signatures vanish completely. This distinction is of considerable importance for the study of optical properties in the vicinity of naked singularities.

To illustrate more directly the influence of the physical parameters on the black hole shadow, we plot the two-dimensional shadow profiles in celestial coordinates $(x/M,y/M)$. Owing to the spherical symmetry of the spacetime, the shadow remains circular in all cases. 
\begin{figure}[htbp]
	\begin{center}
		\subfigure[Varying $Q/M$ for fixed $\ell=0.1$ and $p/M=0$. \label{fig:varyQ}]{
			\includegraphics[width=5cm]{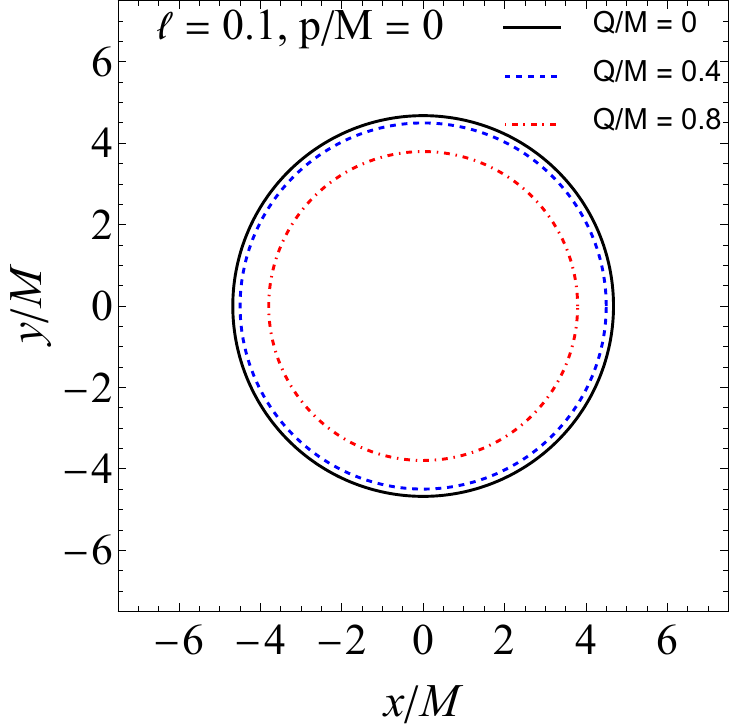}
		}
		\ \
		\subfigure[Varying $p/M$ for fixed $\ell=0.1$ and $Q/M=0$. \label{fig:varyp}]{
			\includegraphics[width=5cm]{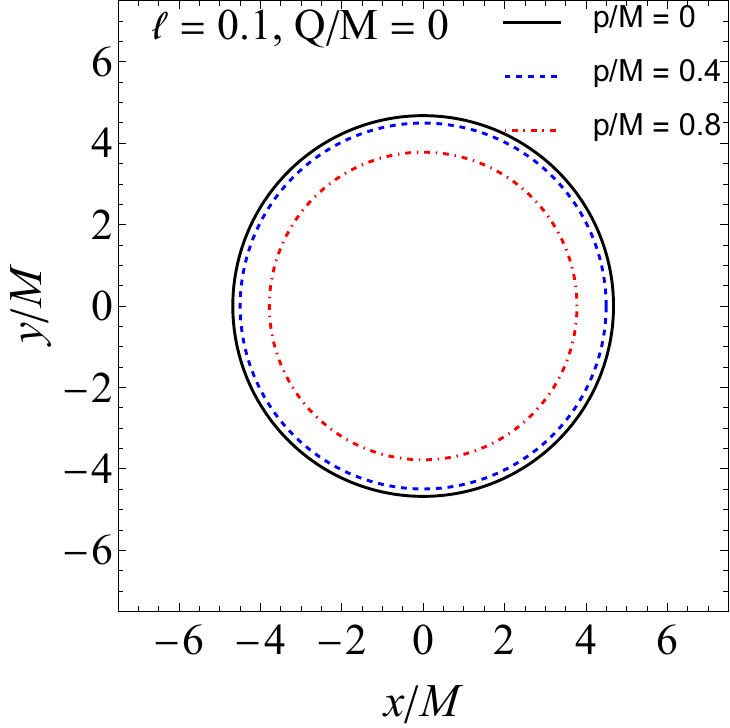}
		}
		\ \
		\subfigure[Varying $\ell$ for fixed $Q/M=0.4$ and $p/M=0.4$. \label{fig:varyell}]{
			\includegraphics[width=5cm]{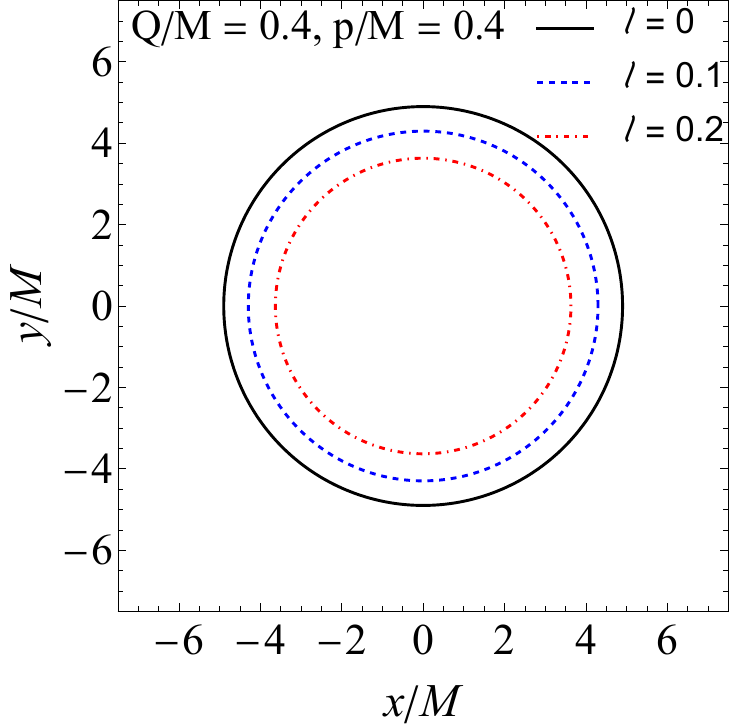}
		}
		\caption{Effects of different physical parameters on the two-dimensional black hole shadow profile.}
		\label{fig:shadow_profiles}
	\end{center}
\end{figure}
As shown in Fig.~\ref{fig:shadow_profiles}, in the presence of Lorentz violation the electric charge $Q/M$ and the magnetic charge $p/M$ have very similar effects on the shadow size: increasing either $Q/M$ or $p/M$ leads to an appreciable decrease in the shadow radius. This indicates that stronger electromagnetic charges weaken the effective gravitational attraction near the horizon, causing the photon sphere to move inward and thus reducing the critical impact parameter for photon capture. On the other hand, when the electromagnetic charges are held fixed, the shadow also shrinks significantly as the Lorentz-violating parameter $\ell$ increases. This shows that Lorentz violation modifies not only the overall spacetime structure of the black hole, but also the bending of light in its vicinity. Such behavior may help distinguish this class of black holes from the Schwarzschild solution in future high-precision observations, for example with the Event Horizon Telescope (EHT).
\subsection{Innermost Stable Circular Orbit of Massive Particles}

For massive test particles ($\epsilon=1$), the effective potential becomes
\begin{equation}
	V_{\text{eff}}(r)=\sqrt{F(r)\left(1+\frac{L^2}{r^2}\right)}.
\end{equation}
If a particle moves on a stable circular orbit, not only must the radial velocity and radial acceleration vanish simultaneously, namely $V_{\text{eff}}=E$ and $\partial_r V_{\text{eff}}=0$, but the orbit must also remain stable under small perturbations, which requires $\partial_r^2 V_{\text{eff}}>0$. From the first-order extremum condition $\partial_r V_{\text{eff}}=0$, one obtains
\begin{equation}
	F'(r)\left(1+\frac{L^2}{r^2}\right)-F(r)\frac{2L^2}{r^3}=0.
\end{equation}
Accordingly, the square of the angular momentum required to sustain a circular orbit at radius $r$ can be written as
\begin{equation}
	L^2=\frac{r^3F'(r)}{2F(r)-rF'(r)}.
	\label{eq:L2}
\end{equation}
Substituting this relation back into the energy condition yields the square of the orbital energy,
\begin{equation}
	E^2=\frac{2F(r)^2}{2F(r)-rF'(r)}.
	\label{eq:E2}
\end{equation}

The innermost stable circular orbit (ISCO) corresponds to the inner boundary of the region where stable circular orbits exist. It is determined by the critical condition under which the minimum of the effective potential degenerates into an inflection point. One therefore further requires
\begin{equation}
	\partial_r^2V_{\text{eff}}=0.
\end{equation}
Expanding the second derivative gives
\begin{equation}
	F''(r)\left(1+\frac{L^2}{r^2}\right)-\frac{4F'(r)L^2}{r^3}+\frac{6F(r)L^2}{r^4}=0.
\end{equation}
Eliminating $L^2$ by substituting Eq.~\eqref{eq:L2} into the above expression, the ISCO condition can be reduced to an equation depending only on the metric function and its derivatives:
\begin{equation}
	r_{\text{ISCO}}
	\left[
	2F'(r_{\text{ISCO}})^2
	-
	F(r_{\text{ISCO}})F''(r_{\text{ISCO}})
	\right]
	-
	3F(r_{\text{ISCO}})F'(r_{\text{ISCO}})
	=0.
	\label{eq:ISCO_condition}
\end{equation}
Substituting Eq.~\eqref{eq:F_full} into Eq.~\eqref{eq:ISCO_condition} and simplifying, one obtains the following sixth-order polynomial equation for $r_{\text{ISCO}}$:
\begin{align}
	&4\left( \frac{Q^2}{(1-\ell)^2} + \frac{p^2}{1-2\ell} \right)^2
	- 9M\left( \frac{Q^2}{(1-\ell)^2} + \frac{p^2}{1-2\ell} \right) r_{\text{ISCO}}
	+ 6M^2 r_{\text{ISCO}}^2 \nonumber \\
	&\quad - \frac{M}{1-\ell} r_{\text{ISCO}}^3
	+ \frac{4\Lambda}{1-\ell}\left( \frac{Q^2}{(1-\ell)^2} + \frac{p^2}{1-2\ell} \right) r_{\text{ISCO}}^4
	- \frac{5M\Lambda}{1-\ell} r_{\text{ISCO}}^5  + \frac{4\Lambda}{3(1-\ell)^2} r_{\text{ISCO}}^6 = 0.
	\label{eq:ISCO_poly}
\end{align}
The largest positive real root corresponds to the ISCO radius for a given set of parameters. Since no general analytical solution exists for such a sixth-order polynomial, the ISCO radius must in general be evaluated numerically.

In the $\Lambda = 0$ case, the above equation reduces to the cubic equation
\begin{equation}
	\frac{M}{1-\ell} r_{\text{ISCO}}^3
	- 6M^2 r_{\text{ISCO}}^2
	+ 9M\left( \frac{Q^2}{(1-\ell)^2} + \frac{p^2}{1-2\ell} \right) r_{\text{ISCO}}
	- 4\left( \frac{Q^2}{(1-\ell)^2} + \frac{p^2}{1-2\ell} \right)^2 = 0.
	\label{eq:ISCO_cubic}
\end{equation}
With the aid of Cardano's formula, its analytical solution can be expressed as
\begin{equation}
	r_{\text{ISCO}}
	=
	2M(1-\ell)
	+\sqrt[3]{-\frac{B}{2}+\sqrt{\Delta}}
	+\sqrt[3]{-\frac{B}{2}-\sqrt{\Delta}},
	\label{eq:r_isco_analytical}
\end{equation}
where
\begin{equation}
	A = 9(1-\ell)\left(\frac{Q^2}{(1-\ell)^2} + \frac{p^2}{1-2\ell}\right) - 12M^2(1-\ell)^2,
	\label{eq:param_p}
\end{equation}
\begin{equation}
	B = -\frac{4(1-\ell)}{M}\left(\frac{Q^2}{(1-\ell)^2} + \frac{p^2}{1-2\ell}\right)^2
	+ 18M(1-\ell)^2\left(\frac{Q^2}{(1-\ell)^2} + \frac{p^2}{1-2\ell}\right)
	- 16M^3(1-\ell)^3,
	\label{eq:param_q}
\end{equation}
\begin{equation}
	\Delta = \left(\frac{B}{2}\right)^2 + \left(\frac{A}{3}\right)^3.
	\label{eq:discriminant}
\end{equation}
In the degenerate limit without electromagnetic charges, $Q=p=0$, the above result reduces to
\begin{equation}
	r_{\text{ISCO}}=6(1-\ell)M.
\end{equation}

\begin{figure}[htbp]
	\begin{center}
		\subfigure[Varying $p/M$ for fixed $Q/M=0.1$. \label{fig:isco_p}]{
			\includegraphics[width=7cm]{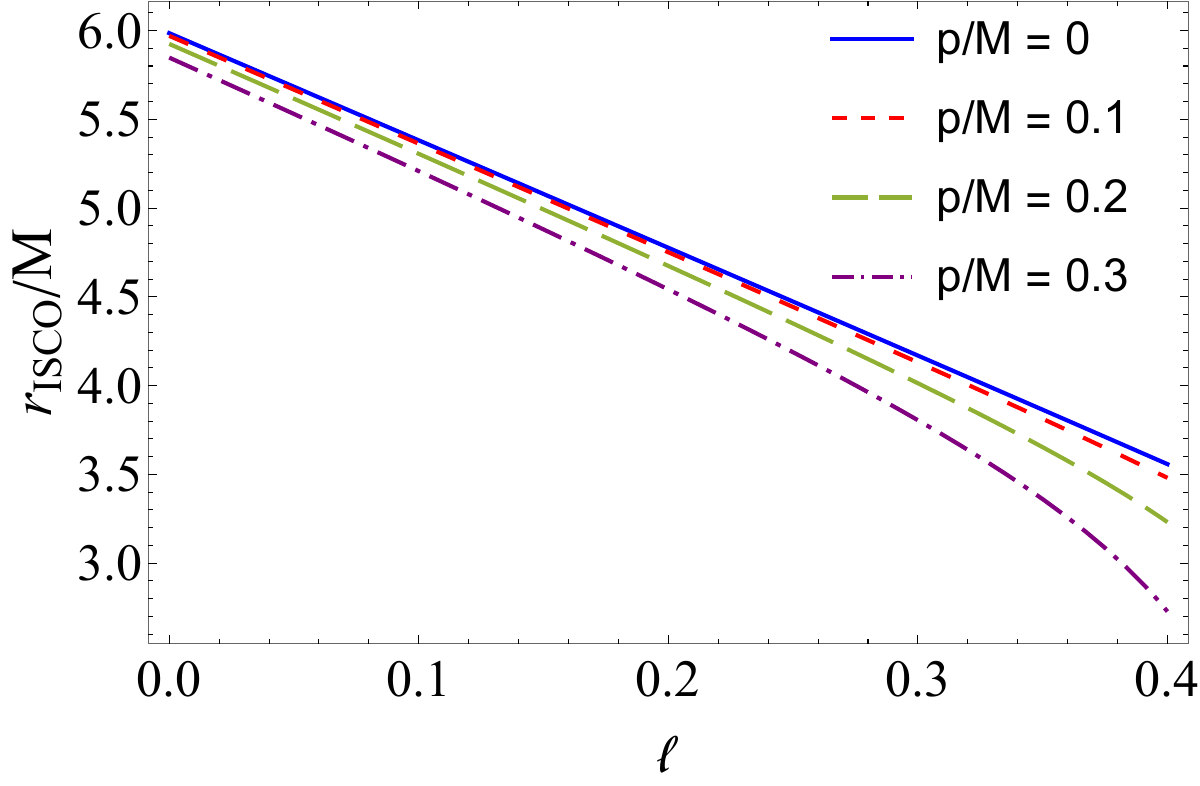}
		}
		\ \
		\subfigure[Varying $Q/M$ for fixed $p/M=0.1$. \label{fig:isco_q}]{
			\includegraphics[width=7cm]{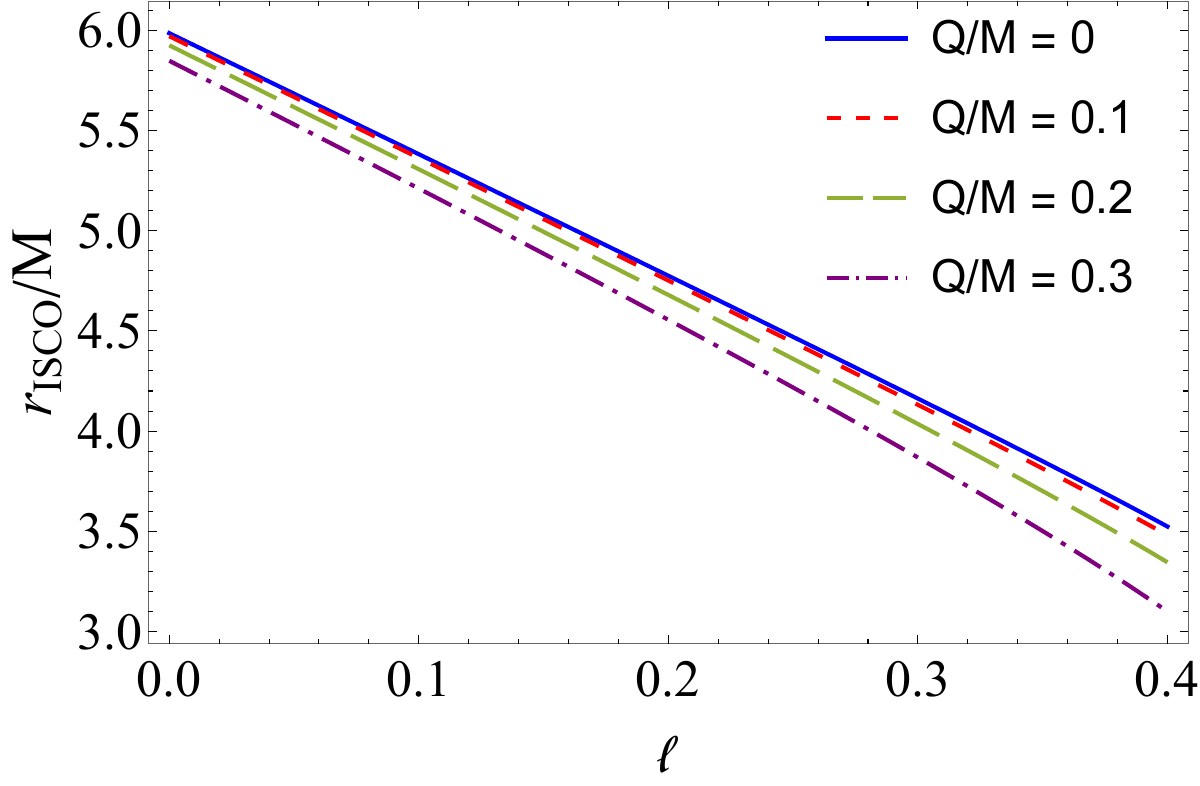}
		}
		\caption{Effects of the Lorentz-violating parameter $\ell$ and the electromagnetic charges on the radius of the innermost stable circular orbit (ISCO).}
		\label{fig:isco_compare}
	\end{center}
\end{figure}

Figure~\ref{fig:isco_compare} illustrates the behavior of the ISCO radius as the Lorentz-violating parameter and the electromagnetic charges are varied. For fixed electromagnetic charges, $r_{\mathrm{ISCO}}/M$ decreases monotonically with increasing $\ell$, indicating that Lorentz-violating effects drive stable circular orbits inward. For a given $\ell$, increasing either the magnetic charge $p/M$ or the electric charge $Q/M$ leads to a further decrease in the ISCO radius, showing that electromagnetic charges also compress the region of stable motion. A comparison between Fig.~\ref{fig:isco_p} and Fig.~\ref{fig:isco_q} further shows that, for relatively large $\ell$, the magnetic charge $p$ has a slightly stronger effect on the ISCO than the electric charge $Q$. This difference can be traced to the fact that the magnetic-charge contribution in the metric grows more rapidly than the electric-charge contribution at larger values of $\ell$. Therefore, when the Lorentz-violating parameter and the electromagnetic charges act simultaneously, the parameter region supporting stable circular orbits is substantially reduced.
\section{Thermodynamics}\label{SS6}

One of the most remarkable features of black holes is that they are not only exact solutions of the gravitational field equations, but can also be regarded as thermodynamic systems obeying a set of fundamental laws~\cite{Bekenstein_1973, Bardeen_1973, Hawking_1975}. In extended phase-space thermodynamics, where the cosmological constant is interpreted as thermodynamic pressure, black holes exhibit not only quantities such as temperature, entropy, and heat capacity, but also rich phase structures and critical behavior analogous to those of van der Waals fluids~\cite{Kastor_2009, Dolan_2011, Cvetic_2011, Kubiznak_2012,Xiao_2025}. Motivated by this perspective, we investigate the thermodynamic properties of the dyonic Kalb--Ramond black hole in an asymptotically AdS background, focusing on its energy, temperature, entropy, pressure, volume, and phase-transition behavior.

Within the framework of gravity nonminimally coupled to the Kalb--Ramond field, the thermodynamic energy and entropy of the black hole system should be derived using the Iyer--Wald formalism~\cite{Wald_1993,Iyer_1994,Iyer_1995,Xiao_2024}. Following this approach, the energy $E$ and entropy $S$ receive corrections controlled by the Lorentz-violating parameter $\ell$ and are given by~\cite{Liu_2025KR,Xiao_2026}
\begin{equation}
	E = (1-\ell)M,
	\label{eq:energy}
\end{equation}
\begin{equation}
	S = (1-\ell)\pi r_+^2 = \frac{1-\ell}{4}\mathcal{A}_+,
	\label{eq:entropy}
\end{equation}
where $r_+$ denotes the event horizon, determined by $F(r_+) = 0$, and $\mathcal{A}_+=4\pi r_+^2$ represents the area of the event horizon. This demonstrates that, as a result of Lorentz symmetry breaking, the entropy of this black hole system deviates from the standard Bekenstein-Hawking area law. Imposing the event horizon condition $F(r_+) = 0$ yields
\begin{equation}
	E=\left(1-\ell\right)M = \frac{r_+}{2} + \frac{Q^2}{2(1-\ell) r_+} + \frac{\left(1-\ell\right)p^2}{2\left(1-2\ell\right)r_+} - \frac{\Lambda r_+^3}{6}.
	\label{eq:mass_horizon}
\end{equation}

In extended phase space thermodynamics, the cosmological constant is interpreted as a thermodynamic pressure~\cite{Dolan_2011}. The thermodynamic pressure should be defined as
\begin{equation}
	P = -\frac{\Lambda}{8\pi}.
	\label{eq:pressure}
\end{equation}
Within this framework, the system energy $E$ is identified with the enthalpy. As a function of entropy $S$, pressure $P$, electric charge $Q$, and magnetic charge $p$, namely $E=E(S,P,Q,p)$, the corresponding first law of thermodynamics reads
\begin{equation}
	dE = \mathcal{T}\, dS + \mathcal{V}\, dP + \Phi_e\, dQ + \Phi_m\, dp,
	\label{eq:first_law}
\end{equation}
where $\mathcal{T}$ is the Hawking temperature, $\mathcal{V}$ is the thermodynamic volume, and $\Phi_e$ and $\Phi_m$ denote the electric and magnetic potentials, respectively.

From the definition of the surface gravity, the Hawking temperature of the dyonic Kalb--Ramond black hole is obtained as
\begin{equation}
	\mathcal{T} = \frac{1}{4\pi}F'(r_+)
	= \frac{1}{4\pi r_+}
	\left[
	\frac{1}{1-\ell}
	- \frac{Q^2}{(1-\ell)^2 r_+^2}
	- \frac{p^2}{(1-2\ell)r_+^2}
	+ \frac{8\pi P r_+^2}{\left(1-\ell\right)}
	\right].
	\label{eq:temperature}
\end{equation}
This leads to the equation of state
\begin{equation}
	P = \frac{\left(1-\ell\right)\mathcal{T}}{2r_+}
	- \frac{1}{8\pi r_+^2}
	+ \frac{Q^2}{8\pi\left(1-\ell\right) r_+^4}
	+ \frac{\left(1-\ell\right)p^2}{8\pi\left(1-2\ell\right)r_+^4}.
	\label{eq:eos}
\end{equation}

For fixed $\ell$, $Q$, and $p$, the critical point of the second-order phase transition is determined by the standard inflection point conditions
\begin{equation}
	\left(\frac{\partial P}{\partial r_+}\right)_{\mathcal{T}} = 0,
	\qquad
	\left(\frac{\partial^2 P}{\partial r_+^2}\right)_{\mathcal{T}} = 0.
	\label{eq:critical_conditions}
\end{equation}
Solving these equations, one obtains the analytical expressions for the critical radius, critical temperature, and critical pressure:
\begin{align}
	r_{c+} &= \sqrt{6(1-\ell)\left(\frac{Q^2}{(1-\ell)^2} + \frac{p^2}{1-2\ell}\right)},
	\label{eq:critical_r} \\
	\mathcal{T}_c &= \frac{1}{3\pi(1-\ell)\sqrt{6(1-\ell)}
		\sqrt{\frac{Q^2}{(1-\ell)^2} + \frac{p^2}{1-2\ell}}},
	\label{eq:critical_T} \\
	P_c &= \frac{1}{96\pi(1-\ell)
		\left(\frac{Q^2}{(1-\ell)^2} + \frac{p^2}{1-2\ell}\right)}.
	\label{eq:critical_P}
\end{align}

From the first law of thermodynamics, the corresponding thermodynamic quantities are derived as
\begin{align}
	\mathcal{V} &= \left(\frac{\partial E}{\partial P}\right)_{S,Q,p}
	= \frac{4\pi }{3}r_+^3, \\
	\Phi_e &= \left(\frac{\partial E}{\partial Q}\right)_{S,P,p}
	= \frac{Q}{\left(1-\ell\right)r_+}, \\
	\Phi_m &= \left(\frac{\partial E}{\partial p}\right)_{S,P,Q}
	= \frac{\left(1-\ell\right)p}{\left(1-2\ell\right) r_+}.
\end{align}
Substituting the above thermodynamic quantities, one finds that the system satisfies the extended Smarr relation
\begin{equation}
	E = 2\mathcal{T}S - 2\mathcal{V}P + \Phi_e Q + \Phi_m p.
\end{equation}

To investigate the local thermodynamic stability of the dyonic Kalb--Ramond black hole, we further compute its heat capacity at constant pressure:
\begin{align}
	C_P &= \mathcal{T}\left(\frac{\partial S}{\partial \mathcal{T}}\right)_P
	= \mathcal{T}\frac{\partial S/\partial r_+}{\partial \mathcal{T}/\partial r_+} \nonumber \\
	&= 2\pi(1-\ell)r_+^2 \frac{ r_+^2 - \frac{Q^2}{1-\ell} - \frac{(1-\ell)p^2}{1-2\ell} +8\pi P r_+^4 }{ -r_+^2 + \frac{3Q^2}{1-\ell} + \frac{3(1-\ell)p^2}{1-2\ell}+8\pi P r_+^4 }.
	\label{eq:heat_capacity}
\end{align}
A positive heat capacity at constant pressure corresponds to a locally thermodynamically stable branch, while divergence points of the heat capacity usually signal a phase transition or a change of stability branch.
\begin{figure}[htbp]
	\begin{center}
		\subfigure[Varying the pressure $P$. \label{fig:Cp_P}]{
			\includegraphics[width=7cm]{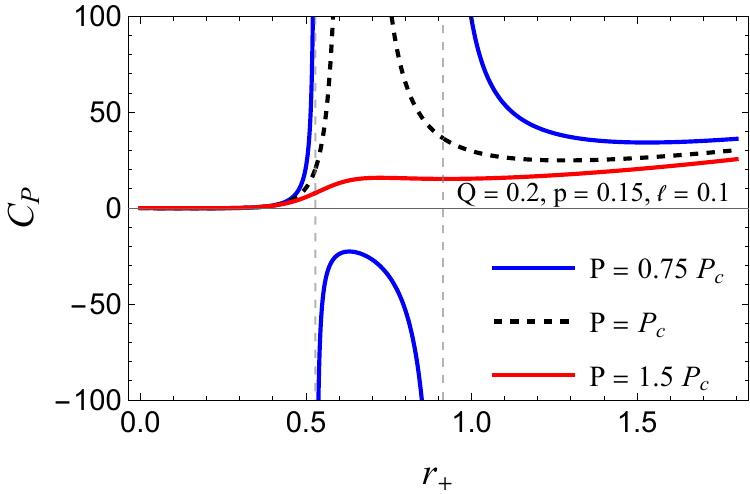}
		}
		\ \
		\subfigure[Varying the magnetic charge $p$. \label{fig:Cp_p}]{
			\includegraphics[width=7cm]{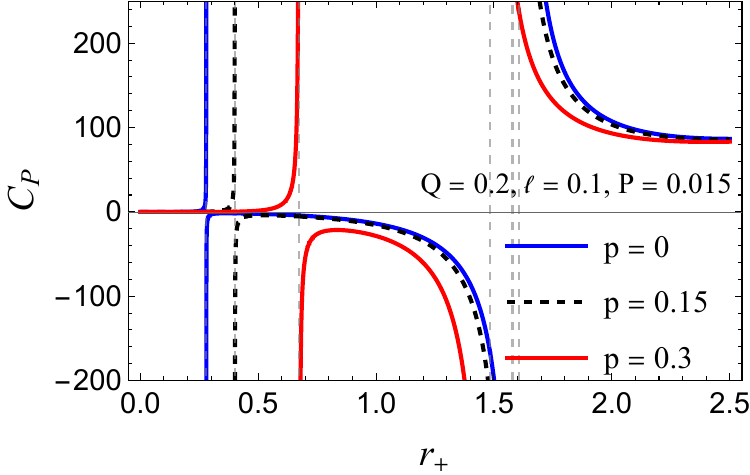}
		}
		
		\subfigure[Varying the electric charge $Q$. \label{fig:Cp_Q}]{
			\includegraphics[width=7cm]{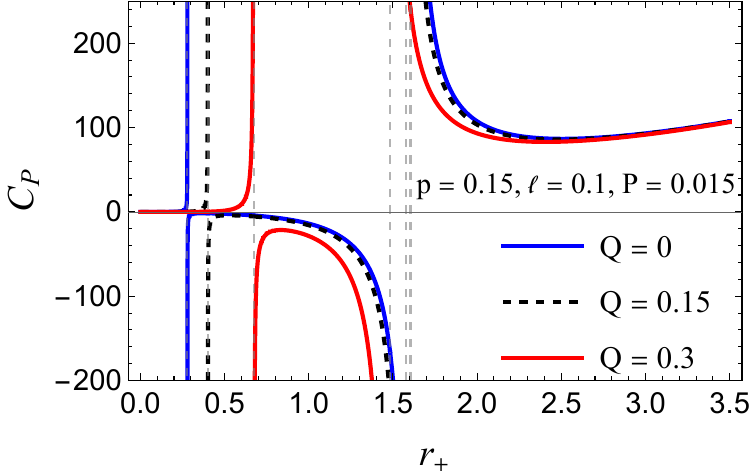}
		}
		\ \
		\subfigure[Varying the Lorentz-violating parameter $\ell$. \label{fig:Cp_ell}]{
			\includegraphics[width=7cm]{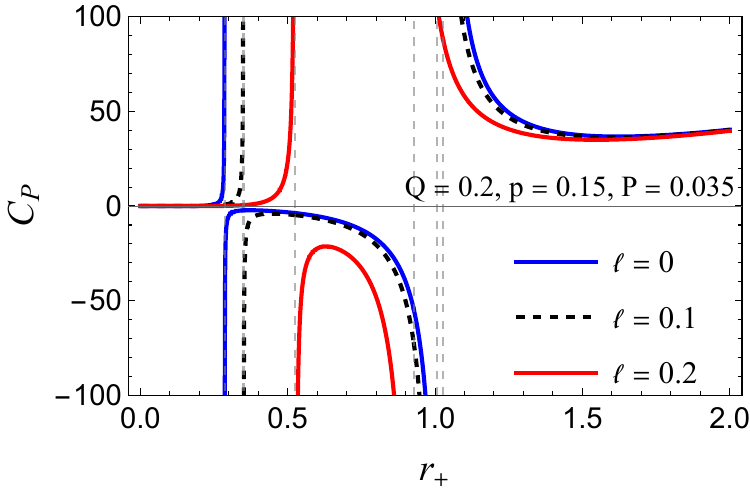}
		}
		\caption{Heat capacity at constant pressure, $C_P$, of the dyonic Kalb-Ramond black hole as a function of the horizon radius $r_+$.}
		\label{fig:Cp_curves}
	\end{center}
\end{figure}
Figure~\ref{fig:Cp_curves} shows that the system exhibits a rich local thermodynamic structure. The pressure $P$ is the key parameter controlling the topology of the phase behavior. When $P<P_c$, the heat-capacity curve develops two Davies divergence points, and a branch of negative heat capacity appears between them, indicating a locally unstable region and a first-order phase transition between small and large black holes. When $P=P_c$, the two divergence points merge into a single critical point, signaling a second-order critical phenomenon. By contrast, when $P>P_c$, the divergence disappears and the system evolves smoothly along a locally stable branch. Variations in the magnetic charge $p$, electric charge $Q$, and Lorentz-violating parameter $\ell$ also shift the locations of the heat-capacity divergences and modify the distribution of stable regions, showing that all of these parameters play an important role in controlling the phase structure of the system.

To further reveal the global thermodynamic stability of the system, we calculate the Gibbs free energy,
\begin{align}
	\mathcal{G} &= E - \mathcal{T}S \nonumber \\
	&= \frac{r_+}{4} + \frac{3Q^2}{4(1-\ell)r_+} + \frac{3(1-\ell)p^2}{4(1-2\ell)r_+} - \frac{2\pi P r_+^3}{3}.
	\label{eq:gibbs}
\end{align}
\begin{figure}[htbp]
	\begin{center}
		\subfigure[Varying the pressure $P$. \label{fig:GT_P}]{
			\includegraphics[width=7cm]{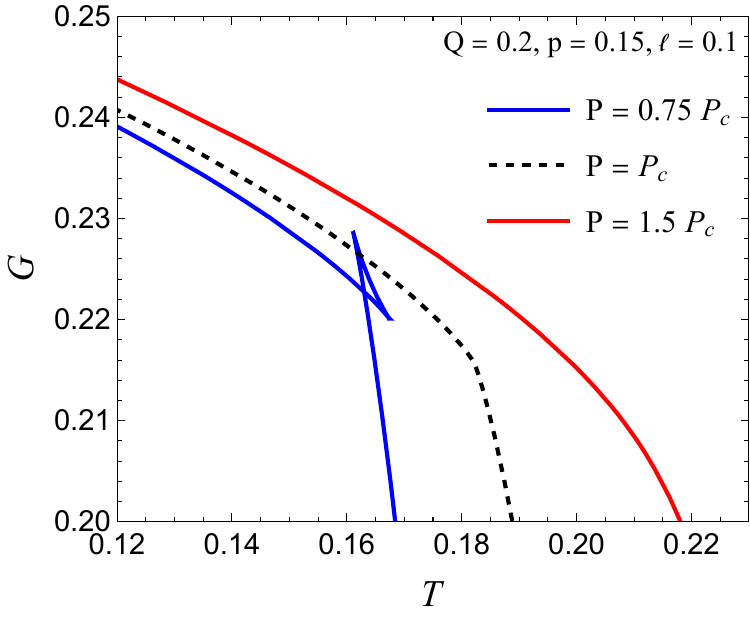}
		}
		\ \
		\subfigure[Varying the magnetic charge $p$. \label{fig:GT_p}]{
			\includegraphics[width=7cm]{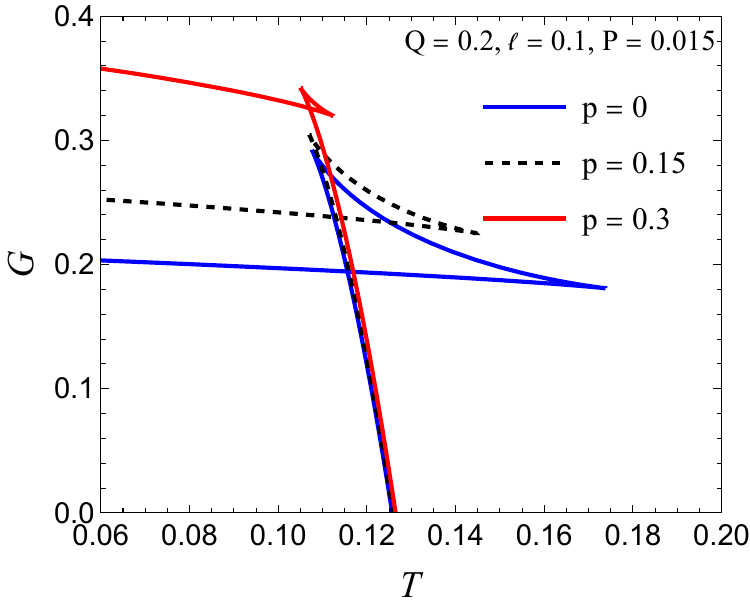}
		}
		
		\subfigure[Varying the electric charge $Q$. \label{fig:GT_Q}]{
			\includegraphics[width=7cm]{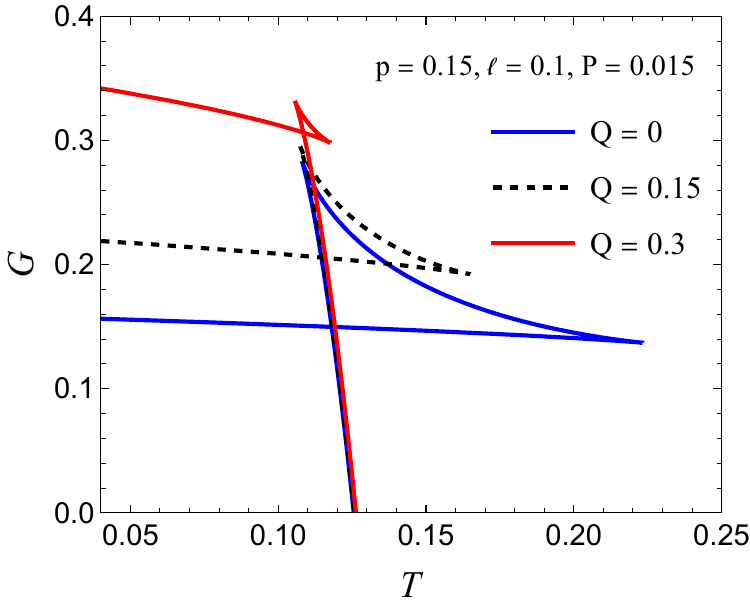}
		}
		\ \
		\subfigure[Varying the Lorentz-violating parameter $\ell$. \label{fig:GT_ell}]{
			\includegraphics[width=7cm]{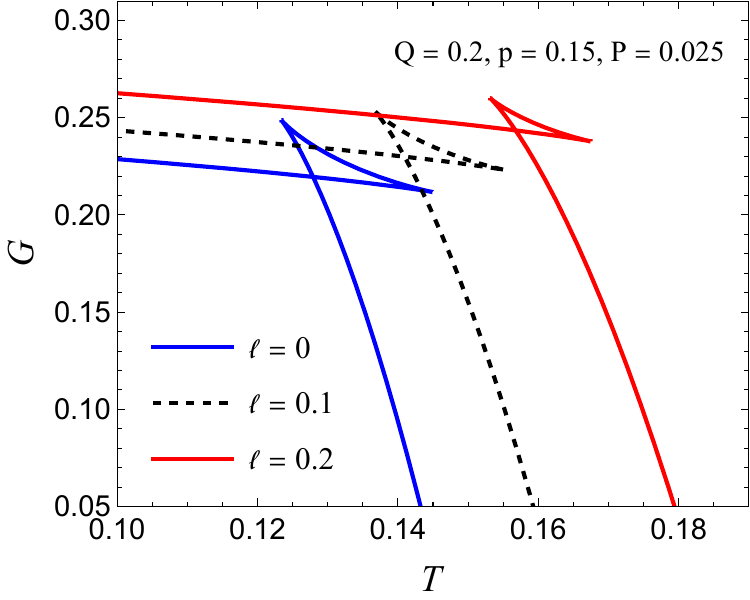}
		}
		\caption{Gibbs free energy $\mathcal{G}$ of the dyonic Kalb--Ramond black hole as a function of the Hawking temperature $\mathcal{T}$.}
		\label{fig:GT_curves}
	\end{center}
\end{figure}
Figure~\ref{fig:GT_curves} displays the typical behavior of the Gibbs free energy as a function of the Hawking temperature. When the pressure lies below the critical value, the free-energy curve develops the characteristic swallowtail structure, which is the standard thermodynamic signature of a first-order phase transition. Because the thermodynamically preferred state is the one with lower free energy, the transition between small and large black holes occurs at the intersection point of the swallowtail branches. As the pressure approaches $P_c$, the swallowtail gradually shrinks to a single critical point, indicating the end of the first-order transition and the onset of second-order critical behavior. When $P>P_c$, the free-energy curve becomes smooth and single-valued, showing that the first-order phase transition disappears completely. This conclusion is fully consistent with the local-stability analysis based on the heat capacity at constant pressure. The magnetic charge $p$, electric charge $Q$, and Lorentz-violating parameter $\ell$ also substantially reshape the swallowtail structure, thereby changing both the temperature interval over which the phase transition occurs and the location of the critical point.

Finally, by extracting the temperature and pressure corresponding to the intersection points of the Gibbs free-energy swallowtail, we construct the coexistence curve between small and large black holes for the dyonic Kalb--Ramond black hole in the $P$--$\mathcal{T}$ plane.
\begin{figure}[htbp]
	\begin{center}
		\subfigure[Varying the Lorentz-violating parameter $\ell$. \label{fig:PT_ell}]{
			\includegraphics[width=7cm]{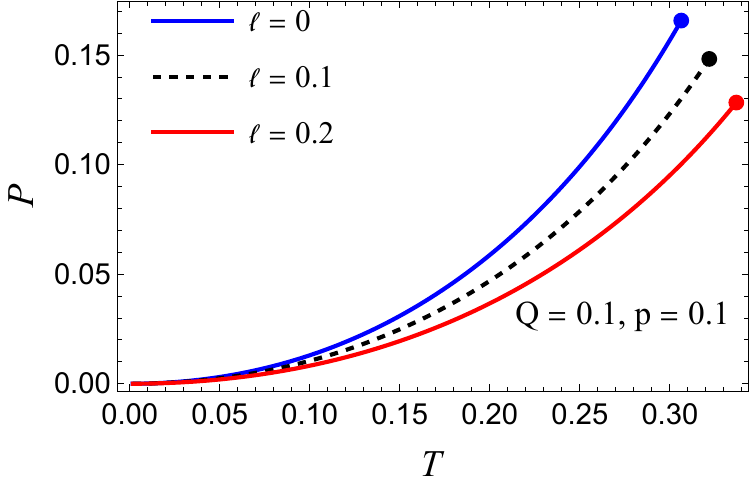}
		}
		\ \
		\subfigure[Varying the magnetic charge $p$. \label{fig:PT_p}]{
			\includegraphics[width=7cm]{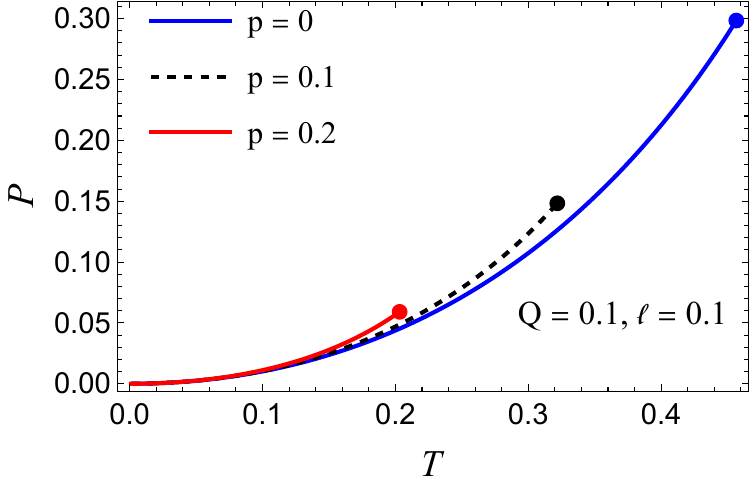}
		}
		
		\subfigure[Varying the electric charge $Q$. \label{fig:PT_Q}]{
			\includegraphics[width=7cm]{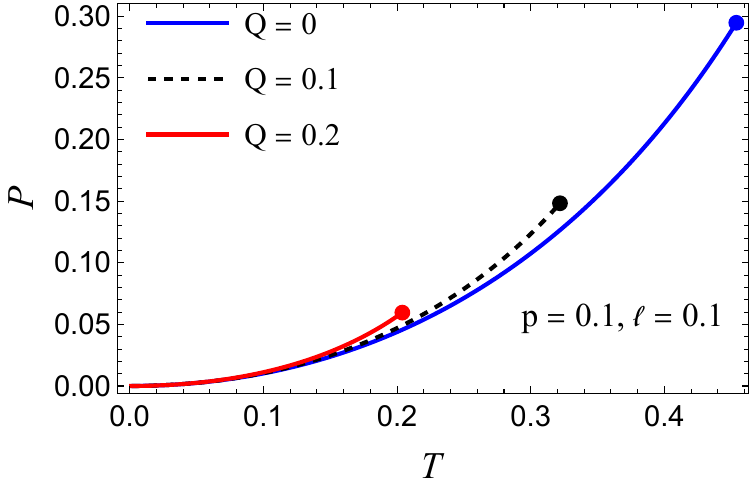}
		}
		\caption{Phase coexistence curves between small black holes and large black holes for the dyonic Kalb-Ramond black hole in the $P$--$\mathcal{T}$ plane. The solid dots at the endpoints of the curves mark the critical points $(\mathcal{T}_c, P_c)$.}
		\label{fig:PT_curves}
	\end{center}
\end{figure}
As shown in Fig.~\ref{fig:PT_curves}, the coexistence curve between small and large black holes starts in the low-temperature, low-pressure region and terminates at the second-order critical point $(\mathcal{T}_c,P_c)$. As the electric charge $Q$, magnetic charge $p$, and Lorentz-violating parameter $\ell$ vary, both the overall position of the coexistence curve and its endpoint shift appreciably, indicating that these parameters not only regulate the domain in which the first-order phase transition takes place, but also significantly influence the critical behavior of the system. Overall, the interplay between Lorentz-violating effects and the electromagnetic dyonic charges endows this black hole system with a thermodynamic phase structure considerably richer than that of the standard RN--AdS black hole.
\section{Conclusions and Outlook}\label{SS7}

In this work, we introduced a nonminimal coupling between the Kalb--Ramond and electromagnetic fields in the presence of both electric and magnetic charges,  and obtained an exact four-dimensional static, spherically symmetric dyonic black hole solution. We have further investigated its geometric properties, orbital dynamics, and thermodynamic behavior in a systematic manner.

From a geometrical perspective, we calculated several curvature invariants for the black-hole spacetime and analyzed their behavior in detail. Our results show that, although the Lorentz-violating parameter can significantly modulate the curvature strength and the divergence structure near the center, it cannot remove the intrinsic physical singularity at the spacetime origin. This indicates that spontaneous Lorentz-symmetry breaking alone is insufficient to resolve the singularity problem. Further improvement of the central geometry will likely require higher-curvature corrections, nonperturbative effects, or a more fundamental quantum-gravity mechanism.

From the perspective of orbital dynamics, we investigated the photon sphere and black hole shadow for massless photons, as well as the   ISCO for massive test particles. The results show that the electric charge, magnetic charge, and Lorentz-violating parameter all induce significant modifications to the effective potential, thereby leading to systematic changes in the photon-sphere radius, the shadow size, and the ISCO radius. In particular, the interplay between the electromagnetic charges and Lorentz-violating effects alters the stability boundaries of particle orbits and may provide a possible astrophysical channel for constraining the Lorentz-violating parameter through observables such as black hole shadows, the inner-edge structure of accretion disks, and related strong-gravity measurements.

On the thermodynamic side, by incorporating the modified energy and entropy derived from the Iyer--Wald formalism and interpreting the cosmological constant as thermodynamic pressure, we established the first law of black hole thermodynamics and the Smarr relation for the dyonic Kalb--Ramond black hole in the extended phase space. We then derived the basic thermodynamic quantities, including the Hawking temperature, entropy, thermodynamic volume, electric potential, and magnetic potential. By analyzing the heat capacity at constant pressure, the Gibbs free energy, and the $P$--$T$ coexistence curve, we found that the system exhibits characteristic van der Waals-type critical behavior, including a first-order phase transition between small and large black holes, a second-order critical point, and phase coexistence. Our results also show that the electric charge, magnetic charge, and Lorentz-violating parameter not only shift the critical point, but also significantly reshape the swallowtail structure of the free energy and the phase-transition trajectory.

Overall, our analysis shows that Lorentz violation induced by the Kalb--Ramond field modifies not only the local and global structure of the black hole spacetime, but also leaves identifiable imprints on particle orbital dynamics and thermodynamic phase transitions. The present model therefore provides a black hole setup with a clear physical interpretation and substantial potential for further investigations of Lorentz-violating effects.

Several directions remain worthy of further study. First, the present model can be extended to slowly rotating, or even fully rotating, configurations in order to examine the influence of angular momentum on the dyonic structure, horizon properties, orbital dynamics, and thermodynamic phase behavior. Second, one may consider more general self-interaction potentials for the Kalb--Ramond field, higher-order nonminimal coupling terms, or additional curvature corrections, so as to investigate the existence and stability of analytical solutions under different vacuum structures, together with possible new phase patterns. Finally, by combining quasinormal modes, gravitational lensing, black hole accretion-disk spectra, shadow observations, and astrophysical data from facilities such as the Event Horizon Telescope, one may hope to place more direct observational constraints on the Lorentz-violating parameter and the electromagnetic charges. Such studies should help clarify the physical implications of black holes in Kalb--Ramond gravity and provide further theoretical support for experimental and observational tests of Lorentz-violating scenarios.
\acknowledgments

This work was supported by the National Natural Science Foundation of China (Grants No. 12475056,  No. 12247101 ), 
the 111 Project (Grant No. B20063), 
Gansu Province's Top Leading Talent Support Plan, the Fundamental Research Funds for the Central Universities (Grant No. lzujbky-2024-jdzx06), and the Natural Science Foundation of Gansu Province (No. 22JR5RA389, No. 25JRRA799). \\
~\\
\textbf{Conflict of Interest:}  The authors declare that they have no conflict of interest.

\bibliographystyle{JHEP}
\bibliography{ref}
\end{document}